\documentclass[sigconf]{acmart}
\AtBeginDocument{%
  }
\copyrightyear{2026}
\acmYear{2026}
\setcopyright{cc}
\setcctype{by}
\acmConference[ICSE-FoSE]{2026 IEEE/ACM 48th International Conference on Software Engineering: Future of Software Engineering }{April 12--18, 2026}{Rio de Janeiro, Brazil}
\acmBooktitle{2026 IEEE/ACM 48th International Conference on Software Engineering: Future of Software Engineering (ICSE-FoSE), April 12--18, 2026, Rio de Janeiro, Brazil}
\acmPrice{}
\acmDOI{10.1145/3793657.3793887}
\acmISBN{979-8-4007-2479-4/2026/04}

\begin{document}
\begin{sloppy}
\title[Rethinking Artifact Evaluation for Software Engineering in the Age of Generative AI]{Rethinking Artifact Evaluation for Software Engineering\\in the Age of Generative AI}
\author{Christoph Treude}
\orcid{0000-0002-6919-2149}
\affiliation{%
  \institution{Singapore Management University}
  \city{Singapore}
  \country{Singapore}}
\author{Christopher M.~Poskitt}
\orcid{0000-0002-9376-2471}
\affiliation{%
  \institution{Singapore Management University}
  \city{Singapore}
  \country{Singapore}}
\author{Rashina Hoda}
\orcid{0000-0001-5147-8096}
\affiliation{%
  \institution{Monash University}
  \city{Melbourne}
  \country{Australia}}
\renewcommand{\shortauthors}{Treude et al.}

\begin{abstract}
Peer review in software engineering research operates under tight time constraints, while generative AI has substantially reduced the human effort required to produce polished research narratives. Reviewer attention is often spent on aspects of submissions such as writing quality or literature positioning that have become relatively less effort-intensive to address, rather than on evaluating the scientific substance of a paper. At the same time, assessing whether methods are implemented correctly, analyses are sound, and claims are supported by evidence remains effort-intensive and dependent on human expertise. In software engineering research, this substance is frequently embodied in artifacts, including code, data, evidence and analysis samples, and experimental infrastructure. In this position paper, we argue that artifact evaluation should be treated as a first-class component of peer review. We frame peer review as an attention allocation problem, examine how generative AI weakens narrative quality as a signal of rigor, and argue that artifact evaluation should play a more prominent role in peer review decisions.
\end{abstract}

\ccsdesc[500]{Software and its engineering}
\ccsdesc[500]{Social and professional topics~Computing profession}

\keywords{Software engineering research, artifact evaluation}
\maketitle

\section{Introduction}

Peer review in software engineering research is under increasing pressure. Submission volumes continue to grow, while reviewer time and attention remain limited. Under these conditions, reviews often emphasize aspects of submissions that are relatively quick to assess, such as writing quality, clarity of motivation, relevance to the venue, checklists of methodological limitations, or overall presentation. These aspects matter, but they are often not where the scientific value of a paper lies.

At the same time, the effort profile of producing research papers has changed. Generative AI tools can now help authors polish prose, restructure introductions, strengthen motivations, identify missing citations, and draft summaries of contributions. Many issues that reviewers routinely flag can therefore be addressed quickly and with relatively little human effort. As a result, the effort of reviewers in identifying narrative shortcomings increasingly yields diminishing returns to scientific progress.

What has not become less effort-intensive is the evaluation of the technical substance. Assessing whether methods are implemented correctly, whether analyses are sound, whether results are robust, and whether claims are supported by evidence still requires domain expertise and careful human judgment. In software engineering research, this substance is frequently embodied in artifacts such as code, datasets, scripts, models, and experimental infrastructure \cite{hermann2020community,liu2024research}. In the case of qualitative research where data and participant privacy considerations limit full transparency, artifacts include shared samples of sanitized evidence and data analysis \cite{hoda2024qual}. 

Peer review can be understood as a problem of attention allocation. Reviewers must distribute their effort across multiple dimensions of research quality, including conceptual novelty, methodological soundness, empirical validity, and clarity of presentation. In practice, aspects of a submission that are immediately visible and less time-intensive to assess tend to receive disproportionate attention. Narrative elements can be evaluated directly from the manuscript and yield concrete feedback, while assessing technical substance, particularly when it is embodied in artifacts, often requires additional setup, contextual understanding, and sustained effort. As a result, artifact-related evaluation is frequently abbreviated, deferred, or omitted altogether \cite{winter2022retrospective}.

This imbalance reflects the structural incentives built into the review process. Review forms, reviewer guidelines, methodology checklists, and limited time budgets implicitly reward feedback that can be efficiently generated from the manuscript itself. Even when reviewers recognize that artifacts encode central parts of a contribution, engaging deeply with them is often neither feasible nor incentivized within a typical review cycle.

Generative AI exacerbates this misalignment by lowering the barrier to producing papers that are coherent, well structured, and plausibly situated in the literature, even when the underlying contribution is weak. Narrative quality thus becomes a less reliable signal of scientific rigor, while evaluating the technical substance that is often embodied in artifacts remains effort-intensive and resistant to automation. As a result, reviewers’ limited attention gravitates toward aspects of submissions that are now faster to produce, such as polished narratives or well-structured positioning. At the same time, aspects that are central to the supporting evidence receive less scrutiny, weakening the connection between claims and evidence~\cite{winter2022retrospective}.

In this position paper, we argue that peer review practices in software engineering have not yet adapted to this shift in the relative value of narrative signals versus evidentiary scrutiny. We first clarify what constitutes scientific evidence across different contribution types, then examine how current artifact evaluation practices fall short of supporting evidence-based review, and finally argue for treating artifact evaluation as a first-class component of peer review in an AI-augmented research ecosystem.

\section{Scientific Evidence in Software Engineering}

Peer review requires reviewers to assess the scientific contribution of a paper, yet what constitutes that contribution is often left implicit. In software engineering research, this ambiguity is amplified by the diversity of contribution types, which differ substantially in how claims are substantiated and what constitutes appropriate evidence. Making these differences explicit is a necessary step toward aligning review practices with the scientific substance of a contribution and toward understanding when artifacts function as primary evidence rather than supplementary material.

We consider the categorization first used in the ICSE 2014 call for papers\footnote{\url{https://2014.icse-conferences.org/research/index.html}} as an example:

\textbf{Analytical}: A paper in which the main contribution relies on new algorithms or mathematical theory. Examples include new bug prediction techniques, model transformations, algorithms for dynamic and static analysis, and reliability analysis. Such a contribution must be evaluated with a convincing analysis of the algorithmic details, whether through a proof, complexity analysis, or run-time analysis, among others and depending on the objectives.

\textbf{Empirical}: A paper in which the main contribution is the empirical study of a software engineering technology or phenomenon. This includes controlled experiments, case studies, and surveys of professionals reporting qualitative or quantitative data and analysis results. Such a contribution will be judged on its study design, appropriateness and correctness of its analysis, and threats to validity. Replications are welcome.

\textbf{Technological}: A paper in which the main contribution is of a technical nature. This includes novel tools, modeling languages, infrastructures, and other technologies. Such a contribution does not necessarily need to be evaluated with humans. However, clear arguments, backed up by evidence as appropriate, must show how and why the technology is beneficial, whether it is in automating or supporting some user task, refining our modeling capabilities, improving some key system property, etc.

\textbf{Methodological}: A paper in which the main contribution is a coherent system of broad principles and practices to interpret or solve a problem. This includes novel requirements elicitation methods, process models, design methods, development approaches, programming paradigms, and other methodologies. The authors should provide convincing arguments, with commensurate experiences, for why a new method is needed and what the benefits of the proposed method are.

\textbf{Perspectives}: A paper in which the main contribution is a novel perspective on the field as a whole, or part thereof. This includes assessments of the current state of the art and achievements, systematic literature reviews, framing of an important problem, forward-looking thought pieces, connections to other disciplines, and historical perspectives. Such a contribution must, in a highly convincing manner, clearly articulate the vision, novelty, and potential impact.

Across these types of contribution, scientific claims are supported through different combinations of conceptual arguments, methodological choices, empirical analyses, and engineering realizations. These dimensions align closely with the long-standing views on experimentation and evidence in software engineering research \cite{wohlin2012experimentation}.

In many contributions to software engineering, particularly empirical and technological papers, engineering realization is inseparable from the methodological and empirical substance. Claims about performance, scalability, usability, security, or developer behavior frequently depend on software systems, datasets, experimental infrastructures, or samples of underlying evidence and analysis procedures that embody design decisions, assumptions, and the `working' of how findings are derived. In such cases, artifacts are not just illustrative; they are primary evidence that determines whether claims are credible \cite{hermann2020community,liu2024research}.

Making explicit what constitutes scientific substance across different contribution types helps surface a recurring mismatch between where evidence resides and where reviewer attention is directed. This clarification provides a foundation for treating artifact evaluation as part of how software engineering research contributions are assessed, rather than as an optional add-on.

\section{State of the Practice in Artifact Evaluation}

Many software engineering venues already support artifact evaluation, reflecting a growing recognition of the importance of artifacts for transparency, reproducibility, and reuse. Early calls to elevate artifacts as part of the scientific evaluation process in software engineering date back at least to the introduction of an Artifact Evaluation Committee at ESEC/FSE 2011, which explicitly argued for treating software and data artifacts as first-class research outputs and reported on one of the first systematic artifact evaluation efforts in the field \cite{krishnamurthi2013artifact}. 

In this section, we use ICSE and the Journal of Systems and Software (JSS) as illustrative examples of current practice, informed by the role of the first author as Open Science Editor for JSS and co-chair of the current ICSE artifact evaluation track. To the best of our knowledge, other major software engineering venues operate in broadly similar ways.

At ICSE, artifact evaluation is organized as a dedicated track that assesses artifacts associated with accepted research papers. Participation is voluntary and occurs after the acceptance decision of the paper. Submitted artifacts are evaluated along various dimensions, including availability, functional consistency with the paper, and reusability, and successful submissions are awarded the corresponding badges. The evaluation focuses on whether the artifacts can be accessed, executed, and understood by the reviewers with reasonable effort and on whether they support the results reported in the paper. The results of the artifact evaluation are published in conjunction with the paper but do not affect acceptance decisions.\footnote{\url{https://conf.researchr.org/track/icse-2026/icse-2026-artifact-evaluation}}

In addition to this post-acceptance process, several conferences have introduced lightweight artifact checks as part of the main review workflow, typically involving a single reviewer verifying the existence and basic accessibility of a repository or accompanying materials. While such checks can help discourage missing or obviously incomplete artifacts, they remain limited in scope and do not constitute a substantive evaluation of artifacts as scientific evidence.

JSS follows a similar post-acceptance model through its Open Science initiative. Authors are encouraged to make the data, code, and other supporting materials publicly available. When authors opt in, these materials are reviewed by the JSS Open Science Board, which evaluates aspects such as accessibility, documentation, and the extent to which the artifacts support the claims made in the article. The outcome of this review is an Open Science statement associated with the accepted paper that signals the availability and quality of the shared materials, while remaining separate from the editorial decision on the manuscript itself.\footnote{\url{https://www.sciencedirect.com/special-issue/10XMGH48FFT}}

Software engineering has published standards and evaluation guidelines that specify how certain types of studies should be designed, conducted, and reported. Examples include the ACM SIGSOFT Empirical Standards, which provide method-specific checklists for experiments, surveys, case studies, and other empirical methods, explicitly intended to guide both authors and reviewers toward consistent judgments of methodological soundness \cite{ralph2020empirical}. Similarly, reporting guidelines for empirical software engineering studies involving large language models emphasize transparency about study design, data collection, analysis procedures, and threats to validity as prerequisites for credible scientific claims \cite{baltes2025guidelines}. Method-specific evaluation guidelines for software engineering studies also exist, e.g., for case studies \cite{runeson2012case}, socio-technical grounded theory \cite{hoda2024qual}, and mixed methods research \cite{storey2025mmr}. 

However, what is largely missing from these standards is a systematic treatment of artifacts as first-class carriers of evidence. Although standards carefully specify what should be reported, they rarely explicitly specify how the underlying software, data, samples of underlying evidence and analyses, and experimental infrastructure should be evaluated when those artifacts embody key methodological decisions or analytic steps. Moreover, qualitative studies can receive even less support in this regard: samples of underlying evidence and illustrations of data analysis procedures are seldom acknowledged as research artefacts in their own right, despite their potential to be presented, evaluated, and to inspire similar studies in the future.

These mechanisms represent a substantial and positive investment in research quality that goes beyond what many venues historically required, and they complement existing methodological and reporting standards by encouraging transparency and reuse. At the same time, their optional nature and post-acceptance placement mean that artifact evaluation remains largely decoupled from judgments about soundness and contribution made during peer review \cite{winter2022retrospective,hermann2020community}. Consequently, artifact evaluation results rarely affect whether claims are considered sufficiently substantiated.

In practice, the scope of artifact evaluation is also limited by feasibility. Reviewers typically focus on whether the artifacts are accessible, documented, and executable in a reasonable time frame. Evaluation often emphasizes installability, basic functionality, and the ability to reproduce reported results using provided scripts or configurations. Although these checks are essential foundations for reuse and reproducibility, and for qualitative work, for transparency through shared evidence samples and analysis traces, they generally do not provide a deeper investigation of whether artifacts adequately support the claims of the article, expose key assumptions, or enable exploration beyond the reported results.

These limitations reflect the difficulty of treating artifacts as scientific evidence under current review conditions. Meaningful artifact evaluation often requires substantial domain expertise, familiarity with specific tools or environments, and time to understand complex pipelines involving code, data (or snippets of underlying evidence), samples of analyses procedures, and infrastructure. Even with containerization, detailed documentation, and reviewer-author interaction, establishing confidence in the evidentiary role of an artifact can be challenging within typical review timelines \cite{winter2022retrospective}.

Structural factors further reinforce this gap. Artifact evaluation is commonly handled by separate committees or boards whose work, while time-intensive, is less visible than that of paper reviewers. Previous studies highlight challenges such as unclear expectations about the depth of the evaluation, mismatches between the intent of the author and the interpretation of the reviewer, and limited incentives for sustained investment in high-quality artifacts \cite{hermann2020community,timperley2021understanding}.

For authors, this arrangement sends mixed signals. Although artifact availability and quality are encouraged and rewarded, the benefits of investing heavily in artifacts are often indirect. This reinforces incentives to prioritize narrative presentation over strengthening the evidentiary basis of a contribution.

\section{Making Artifact Evaluation First-Class}

First-class artifact evaluation means treating artifacts as integral components of a research contribution when scientific claims depend on software, data, or tooling. Rather than being evaluated as optional supplements or post-hoc evidence of reproducibility, such artifacts should be considered part of the material on which judgments about soundness, credibility, and contribution are based.

This does not imply uniform expectations across all papers. Instead, the evaluation effort should be aligned with the evidentiary role artifacts play for a given contribution type. In empirical and technological work, where engineering realizations underpin methodological validity or claimed effects, artifact quality should directly inform reviewers’ assessments. In these cases, artifacts function as primary evidence alongside narrative, not merely as enablers of reuse \cite{winter2022retrospective}.

Elevating artifact evaluation to first-class status raises questions of incentives, effort, and risk. Preparing high-quality artifacts requires substantial time and expertise, yet when artifact evaluation is optional or weakly coupled to acceptance decisions, this investment is difficult for authors to justify. By contrast, improvements in narrative presentation, now often supported by generative AI, have a more predictable and immediate effect on review outcomes. This imbalance reinforces incentives to prioritize polish over strengthening the evidentiary basis of a contribution \cite{timperley2021understanding}.

For reviewers, artifact evaluation similarly involves significant effort, but is often less visible and less formally recognized than paper review. Even when artifacts are carefully assessed, this work is frequently framed as an auxiliary service rather than as a core scholarly evaluation. Previous studies of artifact committees highlight challenges such as limited recognition, unclear expectations about evaluation depth, and ambiguous responsibility, all of which can undermine sustained engagement with artifact review \cite{hermann2020community}.

Several risks must therefore be acknowledged. Evaluating artifacts is time-consuming, and elevating their role without reallocating reviewer attention risks increasing overall review burden. Reviewer capacity is already limited, making it essential that first-class artifact evaluation displaces lower-value review activities, such as extensive narrative critique, rather than simply adding new obligations \cite{winter2022retrospective}. In addition, not all software engineering contributions rely on artifacts to the same degree. Conceptual, theoretical, or perspective papers may depend primarily on argumentation, while empirical and technological papers vary widely in how central artifacts are in substantiating claims \cite{hermann2020community}. Qualitative studies where full disclosure of underlying data or evidence is unreasonable, e.g., due to participant privacy and ethics requirements, rely more heavily on reviewers being able to evaluate the evidence samples provided and to follow the analyses procedure illustrations to discern its credibility and rigor.

There is also a risk of conflating evidentiary value with engineering polish. Well-packaged artifacts with extensive automation or infrastructure may appear more convincing than simpler artifacts that nonetheless adequately support the claims being made. This risk is amplified by generative AI tools, which increasingly make it lower-effort to produce artifacts that look polished, well-documented, and aligned with community expectations, even when their evidentiary value is limited. Mitigating this risk requires focusing evaluation on what an artifact enables reviewers and readers to assess: assumptions, analyses, and links between evidence and claims, rather than on aesthetic or infrastructural qualities \cite{timperley2021understanding}.

It is worth considering how artifact-centered evaluation positions the field for future review practices. Recent work has shown that AI systems are already capable of credibly participating in parts of the academic ecosystem, including producing publishable research narratives and interacting with peer review processes \cite{monperrus2025project}. Although current AI systems are not yet reliable judges of scientific validity, artifacts offer structured, executable, and inspectable representations of methods and evidence. This makes them a more plausible target for future AI-assisted evaluation, such as checking experimental pipelines, probing sensitivity to configuration choices, or verifying consistency between claims and results. Treating artifacts as first-class evidence therefore not only realigns human reviewer attention today but also prepares peer review for a future in which some forms of preliminary technical scrutiny may be partially automated.

\section{Conclusions}

Generative AI is reshaping the effort and attention dynamics of research communication. Producing polished narratives, coherent positioning, and well-structured presentations is becoming faster (despite ongoing sustainability concerns), while the careful evaluation of the technical substance remains slower and dependent on human expertise. In this setting, peer review practices that continue to prioritize narrative quality risk misallocating scarce reviewer attention toward aspects of research that are no longer reliable signals of rigor.

This paper has argued that elevating artifact evaluation to first-class status is a necessary adaptation to this shift. It also expands the definition of artifacts to include samples of underlying evidence and illustrations of data analysis procedures in qualitative studies where full disclosure is unreasonable. In software engineering research, artifacts frequently embody the methodological, empirical, and engineering substance on which claims are based. Treating artifacts as explicit objects of evaluation helps realign peer review with the actual structure of evidence in the field, strengthening the connection between claims and their technical realization.

Importantly, this is not a call for uniform requirements, stricter mandates, or increased burden across all submissions. Different types of contribution are based on different forms of evidence, and artifact evaluation must remain sensitive to the role artifacts play in substantiating claims. The shift advocated here is conceptual rather than procedural: recognizing when artifacts function as primary evidence and ensuring that evaluation practices reflect that reality.

Looking ahead, the question is not whether peer review in software engineering will continue to evolve, but how. In an AI-augmented research ecosystem, maintaining credibility, trust, and cumulative progress requires evaluation practices to adapt to changing effort and attention structures, as well as weakening narrative signals. Treating artifact evaluation as first-class is one concrete step toward an evidence-centered model of software engineering research.


\end{sloppy}
\end{document}